# Vector soliton fission by reflection at nonlinear interfaces

Fangwei Ye, Yaroslav V. Kartashov, and Lluis Torner

*ICFO-Institut de Ciencies Fotoniques, and Universitat Politecnica de Catalunya,*

*Mediterranean Technology Park, 08860 Castelldefels (Barcelona), Spain*

We address the reflection of vector solitons, comprising several components that exhibit multiple field oscillations, at the interface between two nonlinear media. We reveal that reflection causes fission of the input signal into sets of solitons propagating at different angles. We find that the maximum number of solitons that arises upon the fission is given by the number of field oscillations in the highest-order input vector soliton.

Spatial vector solitons form when a proper balance between diffraction, self- and cross-modulation in all light components is achieved [1-9]. In cubic media complex vector solitons made of incoherently coupled fields may be stable provided that the strength of cross-modulation coupling does not exceed that of self-modulation. However, strong perturbations modify their internal structure and can lead to their fission. In this Letter we address this phenomenon and reveal that controllable fission does occur by reflection of the vector solitons at the nonlinear interfaces.

The interaction of radiation with nonlinear interfaces gives rise to a number of phenomena including hysteresis, bistability, and surface wave excitation [10-14]. Reflection of scalar solitons at nonlinear interfaces has been explored experimentally in Refs. [15-18]. Such reflection can cause fission of bound soliton states [19], a process that motivates this study. Reflection becomes especially complex when several fields are present [20,21]. We consider reflection of solitons comprising components with multiple field oscillations and find that such process generates sets of diverging scalar solitons. The maximum number of output solitons is given by the number of field oscillations in the highest-order component and is not equal to the overall number of components, as one may expect.

We address the reflection of vector solitons comprising $N$ mutually incoherent field components at the interface of two cubic media with different refractive indices. The evolution of light beams is described by the system of $N$ coupled nonlinear Schrödinger equations for the dimensionless amplitudes $q_n$:

$$i\frac{\partial q_n}{\partial \xi} = -\frac{1}{2}\frac{\partial^2 q_n}{\partial \eta^2} - \left(\sum_{k=1}^{N}|q_k|^2 + H(\eta)\right)q_n. \qquad (1)$$

The transverse $\eta$ and longitudinal $\xi$ coordinates are scaled to the beam width $r_0$ and the diffraction length $L_{\text{dif}} = kr_0^2$, respectively. The function $H(\eta) = 0$ for $\eta \leq 0$ and $H(\eta) = p$ for $\eta > 0$ describes a refractive index jump at $\eta = 0$. For a beam at wavelength $\lambda = 1.55\,\mu\text{m}$ with $r_0 = 10\,\mu\text{m}$ propagating in a medium with refractive index $n_0 = 1.5$, $p = 100$ corresponds to a refractive index step of the order of $10^{-2}$; for a nonlinear coefficient $n_2 \sim 3 \times 10^{-14}\,\text{cm}^2/\text{W}$, $q \sim 1$ corresponds to a field intensity $\sim 10^{10}\,\text{W/cm}^2$. Such interfaces can be implemented in nematic liquid crystals [18], or they can be made by stacking together different materials with substantially different refractive indices [15].

In the absence of an interface vector soliton solutions of Eq. (1) can be found in the form $q_n(\eta,\xi) = w_n(\eta)\exp(ib_n\xi)$. Such solitons contain at least one nodeless component. Components having equal propagation constants $b_k = b_n$ share similar functional shapes. When $b_k \neq b_n$ vector solitons contain components possessing oscillations (see Figs. 1(a), 1(b) for profiles of two-component solitons). At $b_2 \to 0$ the second component vanishes, while at $b_2 \to b_1$ the soliton transforms into two well-separated vector solitons with two-humped total intensity distribution. In cubic media vector solitons may contain a component with $N$ oscillations only if all lower-order components with $N-1, N-2,...$ oscillations are present, thus such solitons include at least $N$ components. The total number of components may exceed $N$, but then some of $w_n$ have similar shapes.

To study the reflection of multi-component solitons at the interface we solved Eq. (1) with input conditions $q_n|_{\xi=0} = w_n(\eta + \eta_0)\exp(i\alpha_{\text{in}}\eta)$, where $\alpha_{\text{in}}$ is the incident angle, and $\eta_0 \gg 1$ ensures that soliton is launched far enough from the interface at $\eta = 0$. For small incident angles $\alpha_{\text{in}}$ the interface reflects solitons almost completely. With increase of $\alpha_{\text{in}}$ the amount of radiation penetrating into the region $\eta > 0$ increases so that one

may resolve both reflected and transmitted beams in the output pattern. For large enough $\alpha_{\text{in}}$ one observes complete soliton refraction. Such behavior occurs for all values of $p$, but larger $p$ require greater incident angles for occurrence of partial and total refraction. Here we are primarily interested in the regime of complete reflection yielding effective vector soliton fission and set $p = 100$.

The typical dynamics encountered with two-component soliton reflection is shown in Fig. 2. While the fundamental mode keeps its profile and amplitude almost unchanged after reflection, thereby exhibiting quasi-particle behavior, the dipole mode experiences large shape transformations upon reflection. Due to the spatial separation between maxima of $w_2$ component, they arrive at the interface at slightly different distances $\xi$. The right pole of $w_2$ component collides with the interface and bounces back in the vicinity of the point $\eta = 0$, while the second pole changes its propagation direction in the vicinity of the location where it meets the reflected right pole and $w_1$ component. Therefore, the left pole of the $w_2$ component is reflected back at a distance from the interface, in contrast to $w_1$ component. This difference in reflection positions leads to different effective potentials experienced by the corresponding fields, and yields different reflection angles (Fig. 2(c)). Notice that upon reflection, the energy concentrated in the right hump of the $w_2$ component couples partially into its left hump, radiative field, and $w_1$ component, so that $w_2$ component looses its dipole-like input structure and reshapes into a single-hump soliton. Thus, collision with interface results in fast fission of input solitons into set of diverging solitons that are scalar for the chosen set of parameters.

The dependence of the angle $\delta\alpha$ between two scalar solitons emerging upon reflection versus incident angle is shown in Fig. 1(c). For the set of parameters in Fig. 1(c) the effective fission of vector soliton occurs at $\alpha_{\text{in}} \lesssim 1.2$, since for larger values of $\alpha_{\text{in}}$ refraction dominates. Note that refraction of vector solitons at large incident angles typically does not result in their fission, i.e. vector solitons rather keep their internal structure after their pass though the interface. At very small angles, $\alpha_{\text{in}} \lesssim 0.01$, the collision is too weak and also does not lead to vector soliton fission. Surprisingly, we found that in the interval $\alpha_{\text{in}} \in [0.01, 1.2]$ the dependence $\delta\alpha(\alpha_{\text{in}})$ is nonmonotonic. The angle $\delta\alpha$ reaches its minimal value at $\alpha_{\text{in}} \approx 0.05$ but it never vanishes. At $\alpha_{\text{in}} > 0.05$ the splitting angle increases monotonically. Notice that $\delta\alpha$ is a nonmonotonic function of $b_2$ as well. The splitting angle vanishes at $b_2 \to 0$, when $w_2$ component goes to zero and may not affect the soliton dynamics, while at $b_2 \to b_1$, vector solitons transform into two separated and almost independent beams, each of them being reflected with almost equal angles. Thus, the most effective fission occurs for intermediate values of $b_2$.

We have found similar fission scenarios for higher-order vector solitons containing more than two input components. Fission of the three-component soliton of Fig. 3(a) whose higher-order component possesses three oscillations is depicted in Fig. 3(c). This soliton breaks into three scalar fragments, with the most intensive fragment (in $w_1$ component) flying apart at the smallest angle, and the less intense fragment (in $w_3$ component) flying apart at the largest angle with respect to $\xi$ axis. Interestingly, one finds that the intensity redistribution inside each component is similar to that for the soliton of Fig. 2: upon reflection, energy concentrated within each component couples into its left outermost hump, which then gives rise to one of scalar solitons, while the minimal distance between this hump and the interface increases with the order of component.

A central result of this Letter is that the number of spatially separated solitons that may emerge upon fission of the vector complexes is determined by the number of oscillations in the highest-order component, and it does not depend on the overall number of components. This point is illustrated in Figs. 3(b) and 3(d) where we show the profile and splitting of a three-component soliton for which $w_1$ and $w_2$ feature similar shapes. One finds that components possessing similar shapes are always reflected with the similar angles irrespectively of the energy concentrated within each component. Hence, e.g., in Fig. 3(d) fission of the input vector soliton give rise to one two-component vector and one scalar soliton. We checked the validity of this rule by conducting extensive numerical simulations of fission of solitons with up to 10 components, having different symmetries.

Notice that results presented here were obtained for the interface of Kerr media and that interfaces between saturable media exhibit different phenomena. Saturable materials support solitons composed of only nodeless component and component featuring multiple field oscillations, provided that the saturation degree exceeds a critical value [7,9]. Our numerical simulations showed that reflection of such solitons gives rise to several solitons with different internal structures. For example, reflection of soliton having nodeless first and three-humped second component results in appearance of scalar soliton and vector soliton composed of nodeless and dipole components (Fig. 4). The difference with the Kerr case is clearly apparent.

# References without titles

# References with titles

# Figure captions

Figure 1.      Profiles of vector solitons at $b_1 = 3$, $b_2 = 1.2$ (a) and $b_1 = 3$, $b_2 = 2.6$ (b). (c) Splitting angle vs incident angle for vector soliton depicted in panel (a). (d) Splitting angle vs $b_2$ for incident angle $\alpha_{\text{in}} = 0.5$ and $b_1 = 3$. In (c) and (d) $p = 100$.

Figure 2 (color online).      Splitting of two-component vector soliton with $b_1 = 3$, $b_2 = 1.2$ into two scalar solitons at $\alpha_{\text{in}} = 1$ and $p = 100$. Panels (a), (b), and (c) show intensities of first and second components, and total intensity, respectively.

Figure 3 (color online).      Profiles of three-component vector solitons at $b_1 = 2.89$, $b_2 = 1.66$, $b_3 = 0.86$ (a), and $b_1 = 3.4$, $b_2 = 3.4$, $b_3 = 0.64$. Splitting of solitons from panels (a) and (b) is depicted in panels (c) and (d), correspondingly. In both cases $\alpha_{\text{in}} = 0.5$ and $p = 100$. In (c) and (d) the total intensity distribution is shown.

Figure 4 (color online).      Splitting of two-component vector soliton with three-humped second component corresponding to $b_1 = 1.06$, $b_2 = 0.66$, at $\alpha_{\text{in}} = 0.5$ and $p = 100$ in saturable medium with $S = 0.8$. Panels (a), (b), and (c) show first component, second component, and total intensity, respectively.

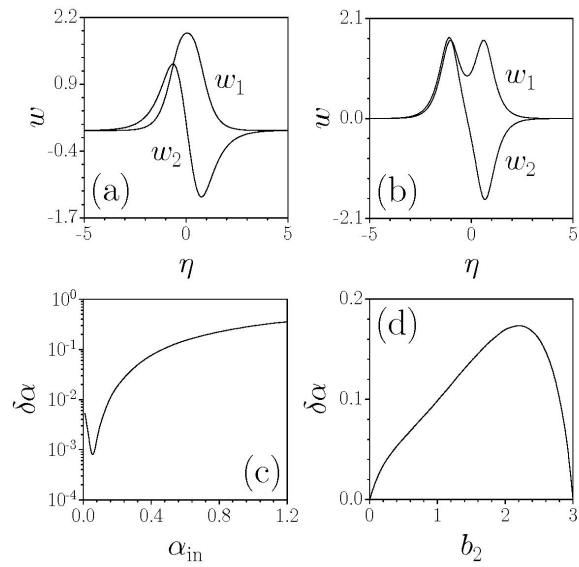

Figure 1. Profiles of vector solitons at $b_1 = 3$, $b_2 = 1.2$ (a) and $b_1 = 3$, $b_2 = 2.6$ (b). (c) Splitting angle vs incident angle for vector soliton depicted in panel (a). (d) Splitting angle vs $b_2$ for incident angle $\alpha_{\rm in} = 0.5$ and $b_1 = 3$. In (c) and (d) $p = 100$.

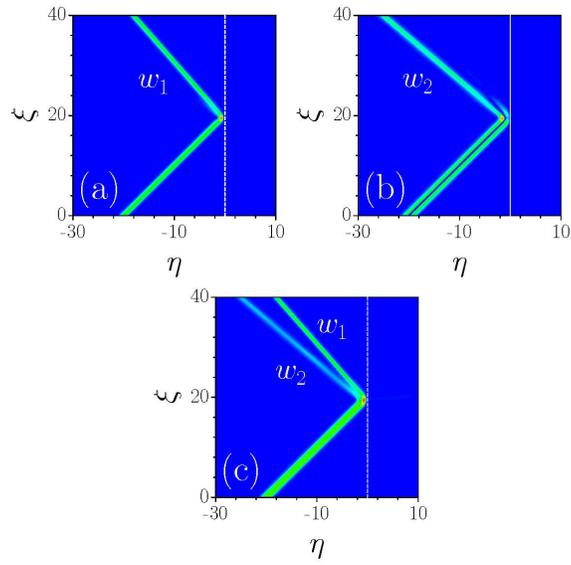

Figure 2 (color online). Splitting of two-component vector soliton with $b_1 = 3$, $b_2 = 1.2$ into two scalar solitons at $\alpha_{\text{in}} = 1$ and $p = 100$. Panels (a), (b), and (c) show intensities of first and second components, and total intensity, respectively.

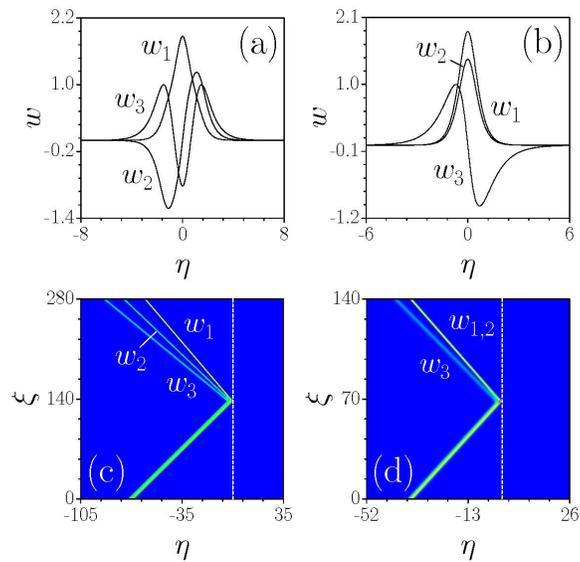

Figure 3 (color online). Profiles of three-component vector solitons at $b_1 = 2.89$, $b_2 = 1.66$, $b_3 = 0.86$ (a), and $b_1 = 3.4$, $b_2 = 3.4$, $b_3 = 0.64$. Splitting of solitons from panels (a) and (b) is depicted in panels (c) and (d), correspondingly. In both cases $\alpha_{\rm in} = 0.5$ and $p = 100$. In (c) and (d) the total intensity distribution is shown.

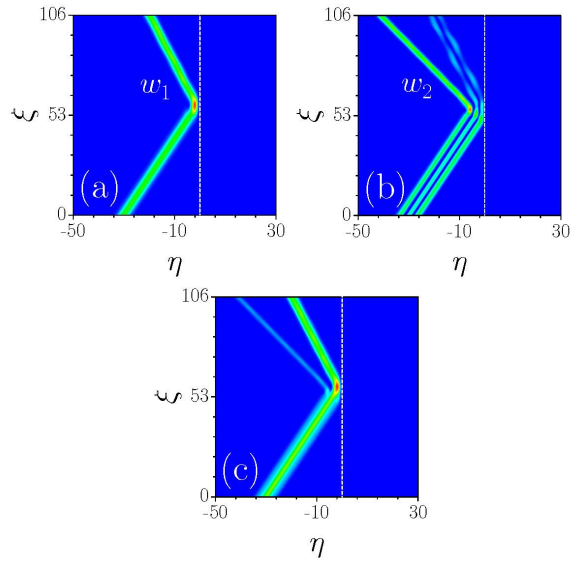

Figure 4 (color online). Splitting of two-component vector soliton with three-humped second component corresponding to $b_1 = 1.06$, $b_2 = 0.66$, at $\alpha_{\text{in}} = 0.5$ and $p = 100$ in saturable medium with $S = 0.8$. Panels (a), (b), and (c) show first component, second component, and total intensity, respectively.